\documentclass[10pt, twocolumn]{IEEEtran}

\usepackage{bm}
\usepackage{url}
\usepackage{bbm}
\usepackage{cite}
\usepackage{tabu}
\usepackage{array}
\usepackage{hhline}
\usepackage{ifthen}
\usepackage{xspace}
\usepackage{dsfont}
\usepackage{balance}
\usepackage{siunitx}
\usepackage{amsmath}
\usepackage{amssymb}
\usepackage{caption}
\usepackage{graphicx}
\usepackage{multicol}
\usepackage{amsfonts}
\usepackage{mathrsfs}
\usepackage{booktabs}
\usepackage{setspace}
\usepackage{hyperref}
\usepackage{makecell}
\usepackage{footnote}
\usepackage{verbatim}
\usepackage{algorithm}
\usepackage{subcaption}
\usepackage{glossaries}
\usepackage{soul, xcolor}
\usepackage[T1]{fontenc}
\usepackage{algpseudocode}
\usepackage{algcompatible}
\usepackage[normalem]{ulem}
\usepackage{multirow, enumitem}

\newcommand{\abs}[1]{\left\lvert#1\right\rvert}

\newcommand{\suchthat}{\;\ifnum\currentgrouptype=16 \middle\fi|\;}

\renewcommand\theadalign{c}
\renewcommand{\tabcolsep}{12pt}
\renewcommand\theadfont{\bfseries}
\renewcommand\cellgape{\Gape[2pt]}
\renewcommand\theadgape{\Gape[2pt]}

\title{Orchestrating UAVs for Prioritized Data Harvesting:\\ A Cross-Layer Optimization Perspective}
\author{Bharath Keshavamurthy~\IEEEmembership{Student Member, IEEE} and Nicol\`{o} Michelusi~\IEEEmembership{Senior Member, IEEE}
\thanks{The source code for this project is available on \href{https://github.com/bharathkeshavamurthy/ACCUSTOM.git}{GitHub}~\cite{Source_Code}.}
\thanks{This work has been supported by NSF under grant CNS-2129015.}
\thanks{The authors are with Electrical, Computer and Energy Engineering,}
\thanks{Arizona State University. Email: \{bkeshav1, nicolo.michelusi\}@asu.edu.}
\vspace{-12.5mm}}

\begin{document}
\bstctlcite{IEEEexample:BSTcontrol}

\maketitle
\thispagestyle{empty}
\pagestyle{empty}
\setulcolor{red}
\setul{red}{2pt}
\setstcolor{red}
\vspace{-13.5mm}

% Abstract
\begin{abstract}
This work describes the orchestration of a fleet of rotary-wing Unmanned Aerial Vehicles (UAVs) for harvesting prioritized traffic from random distributions of heterogeneous users with Multiple Input Multiple Output (MIMO) capabilities. In a finite-horizon offline setting, the goal is to optimize the beam-forming design, the 3D UAV positioning and trajectory solution, and the user association/scheduling policy, to maximize the cumulative fleet-wide reward obtained by satisfying the quality-of-service mandates imposed on each user uplink request, subject to an average per-UAV mobility power constraint. With a probabilistic air-to-ground channel model, a multi-user MIMO uplink communication model with prioritized traffic, and a novel 3D mobility model for rotary-wing UAVs, the 
fleet-wide reward maximization problem is solved via a cross-layer optimization framework: first, K-means clustering is employed to obtain user clusters; then, equipped with a zero-forcing beam-forming design, the positions of the UAVs are optimized via two-stage grid search; next, treating these optimal positions as the graph vertices of a fully-connected mesh, the 3D UAV trajectories (i.e., graph edges) are designed via a learning based competitive swarm optimization algorithm, under an average UAV power consumption constraint, coupled with projected subgradient ascent for dual optimization; consequently, the user association/scheduling strategy is solved via a graphical branch-and-bound method on the underlying multiple traveling salesman problem. Numerical evaluations demonstrate that the proposed solution outperforms static UAV deployments, adaptive Voronoi decomposition techniques, and state-of-the-art iterative fleet control algorithms, vis-\'{a}-vis user quality-of-service and per-UAV average power consumption.
\end{abstract}
\vspace{-1mm}

% Index terms
\vspace{-8mm}

% Introduction, Literature survey, and Novelties
\section{Introduction}\label{S1}
In an increasingly connected world, the widespread adoption of next-generation connectivity technologies---specifically, in the industrial landscape---has led to a multi-fold increase in productivity and yield, while exposing vulnerabilities which when tested by external events (cyber-attacks, disasters) can result in significant harm to economic health and quality of life~\cite{Motivation_1, Motivation_2}. In this regard, within the industrial networking paradigm, this paper envisions the use of Unmanned Aerial Vehicles (UAVs) as data harvesting units to gather critical information from monitoring/aggregator nodes. In particular, the $3$D mobility and maneuverability of UAVs allows us to overcome the limitations of beyond visual line-of-sight links~\cite{MAESTRO_TCCN, SPAVE_ICC}; moreover, autonomous UAVs with intelligent control policies can bypass traditional routes that are rendered inaccessible by external events, and collect priority data for troubleshooting/repair. The primary challenges involved in the orchestration of a fleet of UAVs for data harvesting involves managing the varying degrees of quality-of-service mandates vis-\'{a}-vis the different types of uplink requests generated by the nodes in a typical deployment site, the optimal positioning of the UAVs considering air-to-ground channel characteristics, and the subsequent energy-conscious UAV trajectory design considering their limited on-board energy capabilities.

To tackle these challenges, this paper details an optimization framework to orchestrate MIMO-capable power-constrained rotary-wing UAVs for harvesting prioritized traffic from a random distribution of MIMO-capable heterogeneous users. From a cross-layer optimization perspective, i.e., radio layer + vehicle layer, in an offline finite-horizon centralized setting, we decompose the global fleet-wide reward maximization problem into decoupled sub-problems; consequently, we solve for the optimal positioning of the UAVs and their energy-conscious trajectories, the optimal beam-forming design to maximize MIMO gains, and the user association/scheduling policy.

\noindent{\textbf{Related Work}}: Several papers in the state-of-the-art detail policy frameworks to optimize the orchestration of UAVs in a variety of applications: traffic offloading and coverage extensions for terrestrial base stations using UAV relays~\cite{MAESTRO_TCCN}; maximizing coverage for uplink/downlink communication~\cite{Core_SoA_1_Ref_13, Core_SoA_1_Ref_14}; data harvesting from IoT devices~\cite{Core_SoA_1_Ref_12, Core_SoA_1_Ref_18_Extended_From_17, Core_SoA_1_Ref_25}; computation task offloading to UAV-augmented edge networks~\cite{Core_SoA_1, Core_SoA_1_Ref_24}; capacity-maximizing distributed MIMO backhaul~\cite{CORES_ICASSP, CORES_JSAC}; and wireless power transfer~\cite{Core_SoA_1_Ref_27_Related_To_26}. While these state-of-the-art  works tackle UAV fleet orchestration in non-terrestrial networks, they fail to address practical deployment concerns, particularly in the industrial networking paradigm (Fig.~\ref{F1}). Crucially, unlike the formulation in this paper, the approaches that solve for UAV-assisted data harvesting~\cite{Core_SoA_1_Ref_12, Core_SoA_1_Ref_18_Extended_From_17, Core_SoA_1_Ref_25} fail to model user requests with varied priority levels vis-\'{a}-vis their quality-of-service requirements. Contrary to the optimization perspective presented in this work, in addition to not modeling prioritized traffic, the solutions in~\cite{Core_SoA_1, Core_SoA_1_Ref_24, CORES_ICASSP, CORES_JSAC, Core_SoA_1_Ref_27_Related_To_26} do not account for the on-board energy constraints of the UAVs, while designing their optimal trajectories. Furthermore, the solutions in~\cite{MAESTRO_TCCN, Core_SoA_1_Ref_13, Core_SoA_1_Ref_14} model simplistic and obsolete communication scenarios, i.e., they consider single antenna users and UAVs throughout their constructions; instead, in this paper, we model the use of MIMO-capable users and UAVs, thereby necessitating the need for beam-forming optimization, and yielding spatial multiplexing gains along with multi-user concurrent service.

\noindent{\textbf{Contributions}}: Unlike any other work in the current literature, this paper develops a cross-layer optimization framework based on a model that suitably captures the characteristics of UAV-aided networks in industrial automation environments, i.e., a probabilistic air-to-ground channel model, an uplink multi-user MIMO communication model, and a rotary-wing UAV $3$D mobility power consumption model (with horizontal and vertical accelerations) that involves separating the UAV's $3$D mobility vector into its constituents and accumulating their individual power consumption contributions (Fig.~\ref{F2}). More importantly,
unlike prior work,
in this paper, we account for user requests with varying priority levels, quality-of-service requirements, and commensurate rewards, constituting a variety of traffic flows (Table~\ref{T1}). This model
captures the different types of data offloading requests seen in typical industrial automation deployments (Fig.~\ref{F1}). Additionally, the proposed offline finite-horizon centralized formulation and its resultant cross-layer decomposition, necessitates a solution approach that is unique to UAV-aided networks: a multiple traveling salesman problem formulation to obtain the user association/scheduling solution by employing a graph based branch-and-bound technique. To design the $3$D trajectories of the UAVs, 
subject to an average mobility power constraint per UAV,
we employ a computationally efficient learning-based competitive swarm optimization algorithm and demonstrate its superior convergence performance over state-of-the-art UAV path planning approaches. Finally, our numerical evaluations demonstrate that the proposed framework outperforms static UAV deployments, adaptive Voronoi decomposition schemes, and state-of-the-art iterative fleet control algorithms, vis-\'{a}-vis user quality-of-service and UAV average power consumption.

The rest of this paper is structured as: Sec.~\ref{S2} outlines the system model; Sec.~\ref{S3} elucidates our cross-layer optimization framework along with its constituent algorithms; Sec.~\ref{S4} details our numerical evaluations; Sec.~\ref{S5} lists our conclusions.
\vspace{-8mm}

% System model
\section{System Model}\label{S2}
\begin{figure} [t]
    \centering
    \includegraphics[width=0.84\linewidth]{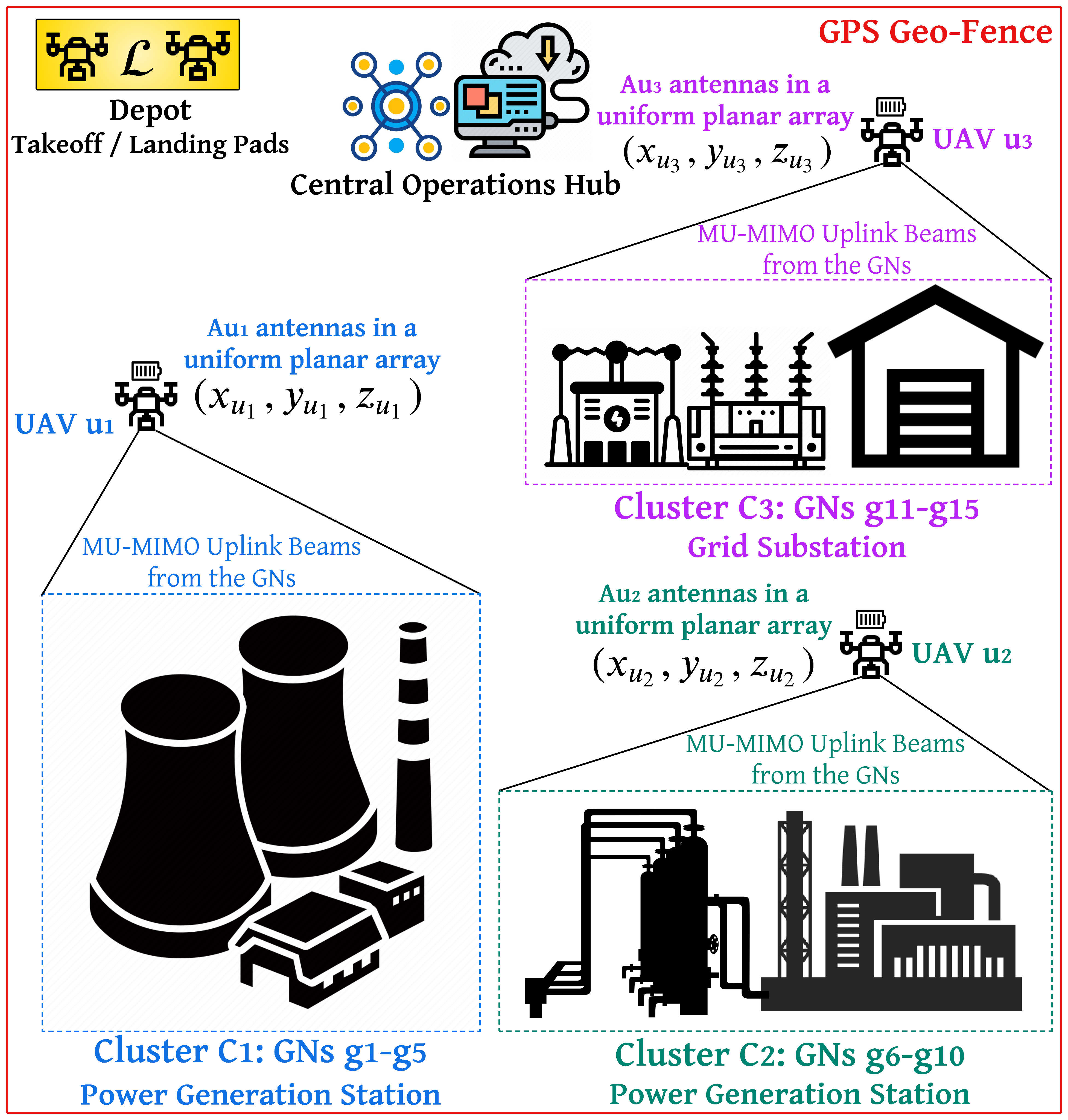}
    \vspace{-1mm}
    \caption{Deployment model for power grid monitoring and restoration.}
    \vspace{-1mm}
    \label{F1}
\end{figure}

In this section, as illustrated in Fig.~\ref{F1}, we detail a model consisting of $U$ MIMO-capable UAVs receiving prioritized traffic from $G$ MIMO-capable users over a finite mission duration. The goal is to find the optimal serving position of each UAV (hover-in-place and receive payload) along with the beam-forming design, design the energy-conscious $3$D UAV trajectories, and derive the user association/scheduling policy.
\noindent{\textbf{Deployment Model}}: As illustrated in Fig.~\ref{F1}, we consider a rectangular deployment site of dimensions $x_{\mathrm{max}}{\times}y_{\mathrm{max}}{\times}z_{\mathrm{max}}$, with the site tessellated into a grid world, where each voxel in this grid is of dimensions $\Delta x{\times}\Delta y{\times}\Delta z$. This deployment site consists of $G$ ground-based users, termed as Ground Nodes (GNs), distributed uniformly throughout---with a GN $g{\in}\mathcal{G}{\triangleq}\{1,2,{\dots},G\}$ equipped with $A_{g}$ antennas arranged in a uniform planar array. With a finite mission duration $T$, the autonomous harvesting of data from these GNs is facilitated by $U$ rotary-wing Unmanned Aerial Vehicles (UAVs)---with a UAV $u{\in}\mathcal{U}{\triangleq}\{1,2,{\dots},U\}$ equipped with $A_{u}$ antennas arranged in a uniform planar array. Note that, to enforce heterogeneity in GN and UAV design, $A_{g_{1}}$ may or may not be equal to $A_{g_{2}}$, for two distinct GNs $g_{1},g_{2}{\in}\mathcal{G},g_{1}{\neq}g_{2}$; and $A_{u_{1}}$ may or may not be equal to $A_{u_{2}}$, for two distinct UAVs $u_{1},u_{2}{\in}\mathcal{U},u_{1}{\neq}u_{2}$. Under a Cartesian coordinate system, we denote the position of a GN $g{\in}\mathcal{G}$ as $\mathbf{p}_{g}{=}(x_{g},y_{g},0)$, where $0{\leq}x_{g}{\leq}x_{\mathrm{max}}$ and $0{\leq}y_{g}{\leq}y_{\mathrm{max}}$. Similarly, at time $t{\in}[0,T]$, the position of a UAV $u{\in}\mathcal{U}$ is denoted by $\mathbf{p}_{u}(t){=}(x_{u},y_{u},z_{u})$, with $0{\leq}x_{u}{\leq}x_{\mathrm{max}}$, $0{\leq}y_{u}{\leq}y_{\mathrm{max}}$, and $0{\leq}z_{u}{\leq}z_{\mathrm{max}}$. Thus, the distance between a GN $g$ and its serving UAV $u$ is described as $d_{ug}{=}\sqrt{(x_{u}{-}x_{g})^{2}{+}(y_{u}{-}y_{g})^{2}{+}z_{u}^{2}}$, and the elevation angle is given by $\theta_{ug}{=}\sin^{-1}{(z_{u}/d_{gu})}$. We define $\mathcal{L}$ to be the set of coordinates corresponding to the takeoff/landing pads for the UAVs such that, for every UAV $u{\in}\mathcal{U}$, $\mathbf{p}_{u}(0){\in}\mathcal{L}$ and $\mathbf{p}_{u}(T){\in}\mathcal{L}$: in other words, we require that all the UAVs start and end their service missions at the depot. Also, to ensure collision avoidance among the UAVs, we enforce the added condition $\mathbf{p}_{u_{1}}(t){\neq}\mathbf{p}_{u_{2}}(t),{\forall}t{\in}[0,T],{\forall}u_{1},u_{2}{\in}\mathcal{U},u_{1}{\neq}u_{2}$. With the grid tessellation simplification outlined earlier, in Sec.~\ref{S3}, this collision avoidance constraint is approached as two or more UAVs not simultaneously occupying the same voxel.\\
\noindent{\textbf{A$2$G Channel Model}}: The channel between a GN $g{\in}\mathcal{G}_{u}$ and its serving UAV $u{\in}\mathcal{U}$ is described by $\mathbf{H}_{ug}{=}\sqrt{\beta}\boldsymbol{\Lambda}\in\mathbb C^{A_u\times A_g}$, where $\beta$ captures the large-scale channel variations while $\boldsymbol{\Lambda}$ captures the small-scale fading effects. We model the large-scale fading component of the channel for both Line-of-Sight (LoS) and Non Line-of-Sight (NLoS) conditions in the  link as
\begin{align}\label{Large_scale}
    \beta_{\mathrm{LoS}}\left(d_{ug}\right) = \beta_{0}d_{ug}^{-\alpha},\ \beta_{\mathrm{NLoS}}\left(d_{ug}\right) = \kappa\beta_{0}d_{ug}^{-\tilde{\alpha}},
\end{align}
where $d_{gu}$ is the distance between GN $g$ and its serving UAV $u$, $\beta_{0}$ denotes the reference pathloss at a distance of $1$ m, $2{\leq}\alpha{\leq}\tilde{\alpha}$ denote the pathloss exponents, and $\kappa$ denotes the additional NLoS attenuation. The LoS probability is given by
\vspace{-3.5mm}
\begin{align}\label{probabilistic}
    \mathbb{P}_{\mathrm{LoS}}\left(\theta_{ug}\right) = \frac{1}{\Big(1 + z_{1}\exp\big\{{-}z_{2}(\theta_{ug} - z_{1})\big\}\Big)},
\end{align}
with the NLoS probability being $\mathbb{P}_{\mathrm{NLoS}}\left(\theta_{ug}\right){=}1{-}\mathbb{P}_{\mathrm{LoS}}\left(\theta_{ug}\right)$, where $z_{1}$ and $z_{2}$ denote propagation parameters specific to the type of radio environment that exists at the deployment site (e.g., rural, urban, suburban), and $\theta_{ug}{\in}(0^{\circ},90^{\circ}]$ is the elevation angle between the GN $g$ and its serving UAV $u$. Next, the distribution of the small-scale fading component $\boldsymbol{\Lambda}$ is also dependent on the LoS/NLoS link state. Specifically, $\boldsymbol{\Lambda}$ is modeled as Rician fading with a $\theta_{ug}$-dependent $K$-factor, i.e., $K\left(\theta_{ug}\right){=}k_{1}\exp\{k_{2}\theta_{ug}\}$ (where $k_{1}$ and $k_{2}$ are environment specific propagation parameters): here, as described in~\cite{Core_SoA_1}, $\boldsymbol{\Lambda}$ constitutes a deterministic term for LoS link states and a term with i.i.d. Rayleigh fading entries for NLoS link states.\\
% Redefining commands
\renewcommand\theadalign{c}
\renewcommand{\tabcolsep}{1.5pt}
\renewcommand\theadfont{\bfseries}
\renewcommand\cellgape{\Gape[1pt]}
\renewcommand\theadgape{\Gape[1pt]}
\begin{table} [tb]
	\centering
    \scriptsize
	\begin{tabular}{|c|c|c|c|c|}
		\hline
		\textbf{Traffic Class} & \textbf{Priority} $\chi$ & \textbf{Max Latency} $\delta_{\mathrm{max}}$ & \textbf{Payload Size} $\nu$ & \textbf{Discount Factor} $\gamma$\\
		\hline
		Telemetry & 100 & 9.1 mins & 256 Mb & 0.10\\
		\hline
        Video & 84 & 11.6 mins & 1387 Mb & 0.24\\
		\hline
        Image & 72 & 14.5 mins & 512 Mb & 0.33\\
		\hline
		File & 24 & 19.0 mins & 536 Mb & 0.80\\
		\hline
	\end{tabular}
    \vspace{-1mm}
	\caption{Quality-of-service table for the network flows in our evaluations.}
	\label{T1}
\end{table}
\noindent{\textbf{Communication Model}}: An uplink transmission request from a GN $g{\in}\mathcal{G}$ is characterized by a request header denoting its traffic class, its assigned priority value $\chi_{g}$, the maximum latency $\delta_{g,\mathrm{max}}$ which defines its quality-of-service constraint, the size of its data payload $\nu_{g}$, and its post-deadline discount factor $\gamma_{g}$ for reward tapering. An example quality-of-service table~\cite{DARPA:SC2} for the variety of prioritized traffic flows considered in our numerical evaluations (Sec.~\ref{S4}) is shown in Table~\ref{T1}. We operate under the assumption that a UAV in the fleet can serve multiple GNs simultaneously, but a GN can only be associated with one UAV, i.e., a GN cannot transmit its data to multiple UAVs. Additionally, once a GN is associated with a UAV, the GN fully uploads its data to the UAV (within a time duration determined by the beam-forming design as well as the channel conditions); upon successfully offloading its data, the GN is considered to have been served by the UAV fleet. We assume that the spectrum allocated to this data harvesting application is discretized into $U$ data channels, each having a preset bandwidth of $B$, with each UAV assigned one of these channels for its service. Let the band-edges of the spectrum be designated as control channels, for inter-UAV communication and for coordination messages between the centralized operations hub and the UAVs. Since the control traffic (UAV-UAV and Hub-UAV) involves very short control frames relative to the large payload frames in the data traffic, we can safely ignore the latencies from control communication in our system model and the subsequent formulations. At time $t{\in}[0,T]$, let the set $\mathcal{G}_{u}{\subseteq}\mathcal{G}$ be defined as the set of GNs associated with UAV $u$. Let $\mathbf{x}_{g}{=}\sqrt{P_{g,\mathrm{Tx}}}\boldsymbol{\Phi}_{g}\mathbf{s}_{g}$ be the signal transmitted to UAV $u{\in}\{1,2,{\dots},U\}$ by GN $g{\in}\mathcal{G}_{u}$, where $P_{g,\mathrm{Tx}}$ denotes the GN's transmit power, $\boldsymbol{\Phi}_{g}{\in}\mathbb{C}^{A_{g}{\times}A_{g}}$ denotes the linear precoding matrix used at the GN, with $\mathrm{tr}(\boldsymbol{\Phi}_{g}\boldsymbol{\Phi}_{g}^H)\leq 1$ and $\mathbf{s}_{g}{\in}\mathbb{C}^{A_{g}{\times}1}$ is the GN's symbol vector with $\mathbb{E}\left[\mathbf{s}_{g}\mathbf{s}_{g}^{H}\right]{=}\mathbf{I}_{A_{g}}$. If $\mathbf{H}_{ug}{\in}\mathbb{C}^{A_{u}{\times}A_{g}}$ is the channel between GN $g$ and UAV $u$, the received signal after combining is
\vspace{0.25mm}
\begin{align}\label{Rx_signal}
    \mathbf{r}_{u} = \boldsymbol{\Gamma}_{gu}\mathbf{H}_{ug}\mathbf{x}_{g} + \boldsymbol{\Gamma}_{gu}\underset{j{\in}\mathcal{G}_{u}{\setminus}g}{\sum}\mathbf{H}_{uj}\mathbf{x}_{j} + \boldsymbol{\Gamma}_{gu}\mathbf{w},
\end{align}
where $\boldsymbol{\Gamma}_{gu}{\in}\mathbb{C}^{A_{g}{\times}A_{u}}$ represents the combining matrix used at the UAV $u$ for GN $g$, and the zero-mean additive white Gaussian noise vector $\mathbf{w}{\sim}\mathcal{CN}(\mathbf{0},BN_{0}\mathbf{I}_{A_{u}})$ after combining becomes $\boldsymbol{\Gamma}_{gu}\mathbf{w}{\sim}\mathcal{CN}(\mathbf{0},BN_{0}\boldsymbol{\Gamma}_{gu}\boldsymbol{\Gamma}_{gu}^{H})$---with $N_{0}$ denoting the power spectral density of the noise and $\mathbf{I}_{A_{u}}$ denoting the identity matrix of size $A_{u}{\times}A_{u}$. Therefore, the MIMO channel capacity (thus, the maximum achievable transmission rate) is
\vspace{-1mm}
\begin{align}\label{Rate}
    &R_{ug}(\beta,\boldsymbol{\Lambda}) = B\log_{2}\det{\left(\boldsymbol{\Upsilon}\right)}\text{, where }\mathbf{H}_{ug}{=}\mathbf{H}_{ug}(\beta,\boldsymbol{\Lambda}),\\
    &\boldsymbol{\Upsilon} = \mathbf{I}_{A_{g}} + P_{g,\mathrm{Tx}} \mathbf{H}_{ug}^{H}\boldsymbol{\Gamma}_{gu}^{H} \mathbf{J}^{-1} \boldsymbol{\Gamma}_{gu}\mathbf{H}_{ug}
    \boldsymbol{\Phi}_{g}\boldsymbol{\Phi}_{g}^H\text{, where}\nonumber\\
    &\mathbf{J}{\triangleq}BN_{0}\boldsymbol{\Gamma}_{gu}\boldsymbol{\Gamma}_{gu}^{H}{+}\boldsymbol{\Gamma}_{gu}\left(\underset{j{\in}\mathcal{G}_{u}{\setminus}g}{\sum}P_{j,\mathrm{Tx}}\mathbf{H}_{uj}\boldsymbol{\Phi}_{j}\boldsymbol{\Phi}_{j}^{H}\mathbf{H}_{uj}^{H}\right)\boldsymbol{\Gamma}_{gu}^{H}.\nonumber
\end{align}
Also, for $\mathbf{J}$ to be invertible, we must have $A_{u}{\geq}A_{g}$. Averaging out the LoS and NLoS link states, we define the average link throughput as $\Bar{R}_{ug}\left(d_{ug},\theta_{ug}\right){=}\mathbb{E}\left[\Bar{R}_{ug}\left(d_{ug},\theta_{ug},\boldsymbol{\Lambda}\right)\right]$, where $\mathbb{E}[{\cdot}]$ is taken over the fading realizations of the channel and
\vspace{-1mm}
\begin{align}\label{throughput}
    \Bar{R}_{ug}\left(d_{ug},\theta_{ug},\boldsymbol{\Lambda}\right) = &\mathbb{P}_{\mathrm{LoS}}\left(\theta_{ug}\right)R_{ug}\Big(\beta_{\mathrm{LoS}}\left(d_{ug}\right),\boldsymbol{\Lambda}\Big) + \nonumber\\&\mathbb{P}_{\mathrm{NLoS}}\left(\theta_{ug}\right)R_{ug}\Big(\beta_{\mathrm{NLoS}}\left(d_{ug}\right),\boldsymbol{\Lambda}\Big).
\end{align}
With this communication model and the GN traffic model described earlier, the reward formulation for the serving UAV is discussed next. Let $\delta_{ug}$ be the time taken by UAV $u$ to harvest the data from GN $g$ (depending on GN-UAV positions, the beam-forming design, and the channel conditions); then, the reward $\Omega_{ug}$ obtained by the UAV is described as follows:
\begin{align}\label{Reward}
    \Omega_{ug} = \chi_{g}\gamma_{g}^{(\delta_{ug} - \delta_{\mathrm{g,max}})}\text{ with }\delta_{ug} = \frac{\nu_{g}}{\Bar{R}_{ug}\left(d_{ug},\theta_{ug}\right)},
\end{align}
i.e., given the discount factor $\gamma_{g}{<}1$, if the UAV $u$ harvests the payload from GN $g$ before the service deadline ($\delta_{\mathrm{max}}$), it gets a higher reward compared to when the UAV goes beyond this service deadline to harvest the GN payload.
\begin{figure} [t]
    \centering
    \includegraphics[width=0.74\linewidth]{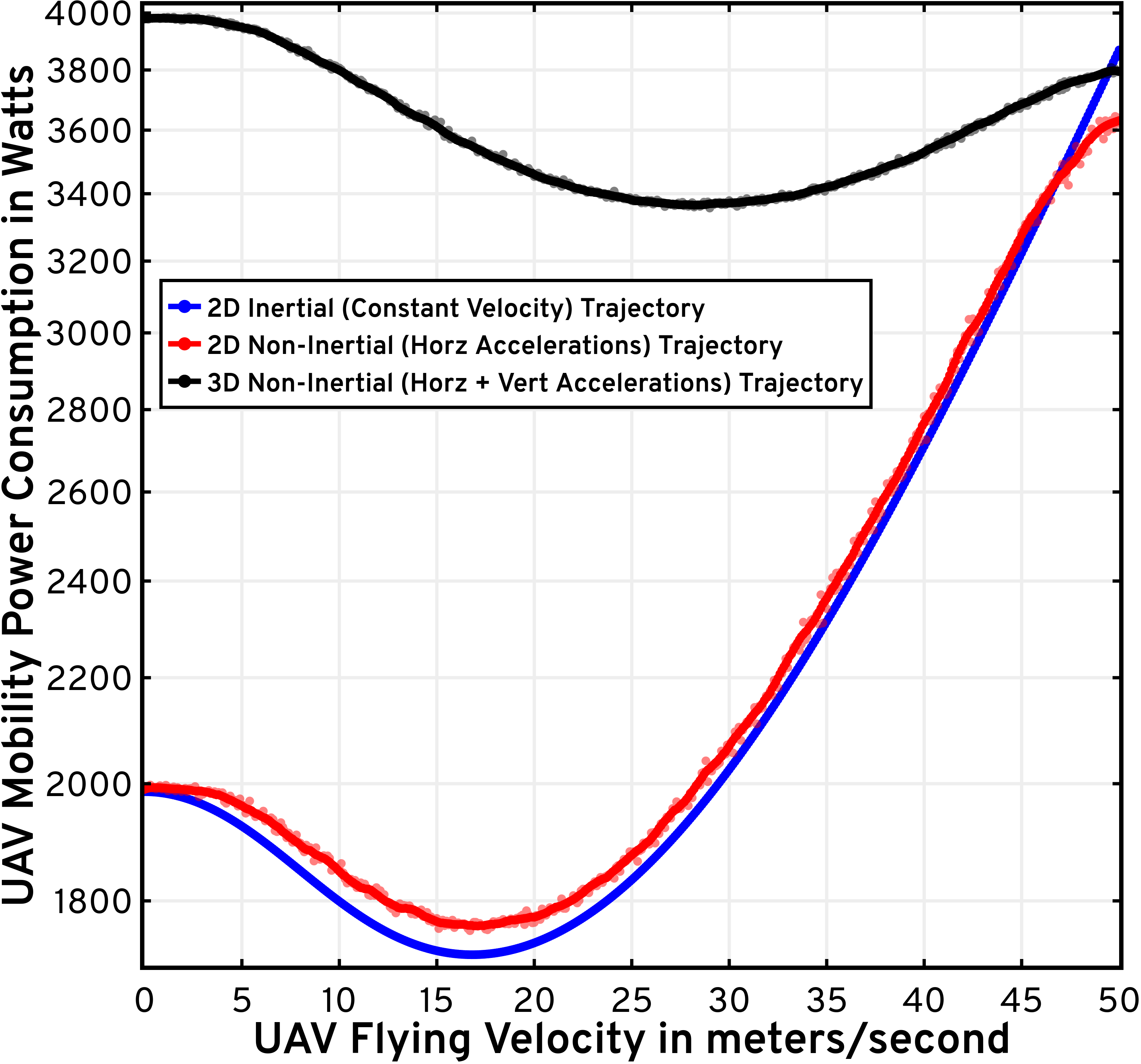}
    \vspace{-1mm}
    \caption{UAV power analyses for $2$D inertial trajectory (blue), $2$D non-inertial trajectory (horizontal accelerations) where its average velocity is equal to the abscissa (red), and $3$D non-inertial trajectory (horizontal and vertical accelerations) where its average velocity is equal to the abscissa (black).}
    \vspace{-1mm}
    \label{F2}
\end{figure}
\noindent{\textbf{UAV Power Model}}: Highlighting the need to accurately model the mobility power consumption of rotary-wing UAVs, Fig.~\ref{F2} depicts the inaccuracies seen in $2$D constant velocity models~\cite{UAV_Propulsion_2, MAESTRO_TCCN}: accounting for vertical propulsion in $3$D motion introduces significant power costs, while introducing horizontal and vertical accelerations results in additional power consumption contributions. In this paper, generalizing the UAV experiments in~\cite{UAV_Propulsion_1}, we aim to alleviate the drawbacks of the widely-used $2$D constant velocity models, by employing vector separation techniques to split the UAV's $3$D motion vector in any given arbitrary trajectory into its constituent horizontal and vertical components, and accumulating their individual power consumption contributions. We define the energy-conscious $3$D trajectory of a UAV $u{\in}\mathcal{U}$ (designed in Sec.~\ref{S3}) as $\mathcal{Q}_{u}{\triangleq}\Big\{\mathbf{p}_{u}(\tau),\vec{v}_{u}(\tau): \tau{\in}\left[t_{u,i},t_{u,f}\right]\Big\}$, where $t_{u,i}$ and $t_{u,f}$ denote the trajectory start and end times, $\mathbf{p}_{u}(\tau)$ denotes the trajectory waypoint at time $\tau$, and $\vec{v}_{u}(\tau)$ is the $3$D velocity vector at time $\tau$, which is separated into its constituent horizontal and vertical components as follows: ${\forall}\tau{\in}\left[t_{u,i},t_{u,f}\right]$,
\vspace{-4.75mm}
\begin{align*}
    v_{u,\mathrm{h}}(\tau){=}\abs{\vec{v}_{u}(\tau)}\cos{\angle{\vec{v}_{u}(\tau)}},\ v_{u,\mathrm{v}}(\tau){=}\abs{\vec{v}_{u}(\tau)}\sin{\angle{\vec{v}_{u}(\tau)}}.
\end{align*}
Therefore, the overall mobility power consumption of a UAV $u{\in}\mathcal{U}$ upon executing a given trajectory is described as follows:
\begin{align}\label{3D_power}
    &P_{u,3\mathrm{D}}\left(\mathcal{Q}_{u}\Big(t_{u,i},t_{u,f}\right)\Big) = \\&\frac{1}{t_{u,\Delta}}\Bigg[E_{u,\mathrm{h}}\left(\Big\{v_{u,\mathrm{h}}(\tau)\Big\}_{\tau{=}t_{u,i}}^{\tau{=}t_{u,f}}\right){+}E_{u,\mathrm{v}}\left(\Big\{v_{u,\mathrm{v}}(\tau)\Big\}_{\tau{=}t_{u,i}}^{\tau{=}t_{u,f}}\right)\Bigg],\nonumber
\end{align}
where $t_{u,\Delta}{=}t_{u,f}{-}t_{u,i}$ is the execution duration of trajectory $\mathcal{Q}_{u}$, $E_{u,\mathrm{h}}({\cdot})$ denotes the energy consumption contributions due to arbitrary accelerating horizontal motion and is given as~\cite{UAV_Propulsion_1}
\begin{align}\label{Horizontal_energy}
    &E_{u,\mathrm{h}}\left(\Big\{v_{u,\mathrm{h}}(\tau)\Big\}_{\tau{=}t_{u,i}}^{\tau{=}t_{u,f}}\right) = \int_{t_{u,i}}^{t_{u,f}}C_{0}\Big(1 + C_{1}v_{u,\mathrm{h}}^{2}(\tau)\Big)d\tau\nonumber\\&{+}\int_{t_{u,i}}^{t_{u,f}}\kappa_{u,\mathrm{h}}(\tau)C_{2}\left(\sqrt{\kappa_{u,\mathrm{h}}^{2}(\tau){+}\frac{v_{u,\mathrm{h}}^{4}(\tau)}{C_{3}^{2}}}{-}\frac{v_{u,\mathrm{h}}^{2}(\tau)}{C_{3}}\right)^{\frac{1}{2}}d\tau\nonumber\\&{+}\int_{t_{u,i}}^{t_{u,f}}C_{4}v_{u,\mathrm{h}}^{3}(\tau)d\tau + \frac{\varrho}{2g}\Big(v_{u,\mathrm{h}}^{2}(t_{u,f}) - v_{u,\mathrm{h}}^{2}(t_{u,i})\Big),
\end{align}
while $E_{u,\mathrm{v}}({\cdot})$ denotes the energy consumption contributions from arbitrary accelerating vertical motion and is given as~\cite{UAV_Propulsion_1}
\begin{align}\label{Vertical_energy}
    &E_{\mathrm{v}}\left(\Big\{v_{u,\mathrm{v}}(\tau)\Big\}_{\tau{=}t_{u,i}}^{\tau{=}t_{u,f}}\right){=}\int_{t_{u,i}}^{t_{u,f}}C_{0}\Big(1{+}C_{1}v_{u,\mathrm{v}}^{2}(\tau)\Big)d\tau\\&{+}\int_{t_{u,i}}^{t_{u,f}}\kappa_{u,\mathrm{v}}(\tau)C_{2}\left(\sqrt{\kappa_{u,\mathrm{v}}^{2}(\tau){+}\frac{v_{u,\mathrm{v}}^{4}(\tau)}{C_{3}^{2}}}{-}\frac{v_{u,\mathrm{v}}^{2}(\tau)}{C_{3}}\right)^{\frac{1}{2}}d\tau\nonumber,
\end{align}
where $a_{u,\mathrm{h}}(\tau){=}\frac{dv_{u,\mathrm{h}}(\tau)}{d\tau}$ is the horizontal acceleration of the UAV at time $\tau$, $a_{u,\mathrm{v}}(\tau){=}\frac{dv_{u,\mathrm{v}}(\tau)}{d\tau}$ is the vertical acceleration of the UAV at time $\tau$, $\kappa_{u,\mathrm{h}}(\tau){=}\kappa\left(v_{u,\mathrm{h}}(\tau),a_{u,\mathrm{h}}(\tau)\right)$ is the UAV's thrust-to-weight ratio for the horizontal plane, and $\kappa_{u,\mathrm{v}}(\tau){=}\kappa\left(v_{u,\mathrm{v}}(\tau),a_{u,\mathrm{v}}(\tau)\right)$ is the UAV's thrust-to-weight ratio for the vertical plane. We define this thrust-to-weight ratio term for a generic velocity and acceleration as follows:
\vspace{-1mm}
\begin{align}\label{Kappa_term}
    \kappa\Big(v(\tau),a(\tau)\Big) = \sqrt{1 + \frac{\Big(\rho\omega\varphi\vartheta v^{2}(\tau) + \frac{2\varrho a(\tau)}{g}\Big)^{2}}{4\varrho^{2}}}.
\end{align}
Here, the site constants $g$ and $\rho$; the UAV design parameters $\varrho$, $\omega$, $\varphi$, and $\vartheta$; and the UAV operational specifications $C_{0}$, $C_{1}$, $C_{2}$, $C_{3}$, and $C_{4}$, are derived from the experiments in~\cite{UAV_Propulsion_1} and are listed in Table~\ref{T2}. Note that, since the UAVs in our formulation are only receiving traffic from the GNs and are not involved in any multi-antenna data transmissions, we can safely ignore their communication power contributions (in the order of $1{-}10$ W, insignificant relative to their mobility power contributions, which are in the order of thousands of watts).
\vspace{-3mm}

% Cross-layer optimization: The radio and vehicle layers
\section{Cross-Layer Optimization Formulation}\label{S3}
\begin{figure}[t]
     \centering
     \includegraphics[width=0.9\linewidth]{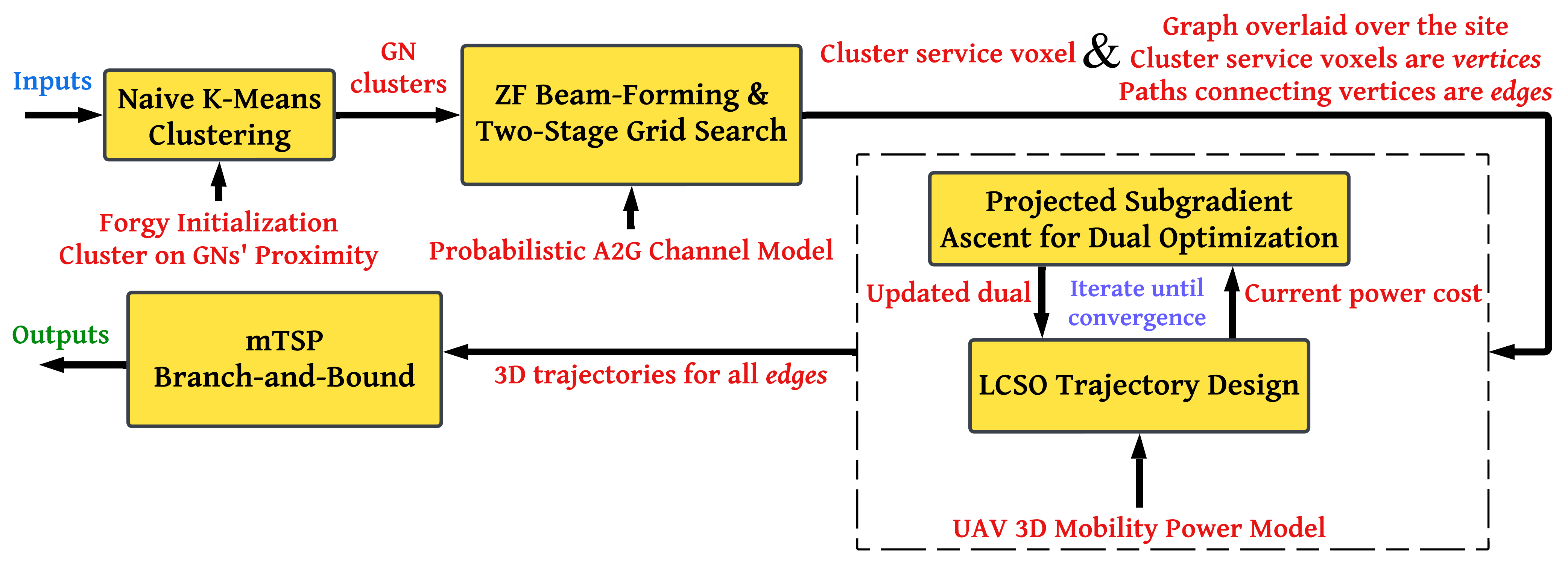}
     \vspace{-2mm}
     \caption{The solution flow of our cross-layer optimization approach.}
     \vspace{-3mm}
     \label{F3}
\end{figure}
% Redefining commands
\renewcommand\theadalign{c}
\renewcommand{\tabcolsep}{3pt}
\renewcommand\theadfont{\bfseries}
\renewcommand\cellgape{\Gape[1pt]}
\renewcommand\theadgape{\Gape[1pt]}
\begin{table} [tb]
    \centering
    \tiny
    \begin{center}
        \begin{tabular}{|*{3}{c|}}
        \hline
        \thead{\textbf{Notation}} & \thead{\textbf{Description}} & \thead{\textbf{Simulation Value}}\\
        \hline
        $T$ & Max mission duration & $3000$ s\\
        \hline
        $U$; $G$ & Number of UAVs; Number of GNs & $6$; $36$\\
        \hline
        $\Delta_{x}$; $\Delta_{y}$; $\Delta_{z}$ & Grid voxel dimensions & $10$ m; $10$ m; $10$ m\\
        \hline
        $x_{\mathrm{max}}$; $y_{\mathrm{max}}$; $z_{\mathrm{max}}$ & Max site dimensions & $3$ km; $3$ km; $150$ m\\
        \hline
        \hline
        $\beta_{0}$ & Reference SNR at $1$ m & $40$ dB\\
        \hline
        $A_{u}$; $A_{g}$ & UAV antenna count; GN antenna count & $16$; $4$\\
        \hline
        $B$; $P_{g,\mathrm{Tx}}$ & Channel bandwidth; GN transmission power & $5$ MHz; $23$ dBm\\
        \hline
        $\alpha$, $\tilde{\alpha}$; $\kappa$ & LoS, NLoS pathloss exponents; NLoS attenuation & $2$, $2.8$; $0.2$\\
        \hline
        $z_{1}$, $z_{2}$; $k_{1}$, $k_{2}$ & LoS probability parameters; Rician $K$-factor parameters & $9.61$, $0.16$; $1$, $0.05$\\
        \hline
        \hline
        $C_{0}$ & UAV power constant vis-\'{a}-vis blade profile & $1276.46$ W\\
        \hline
        $C_{1}$ & UAV power constant vis-\'{a}-vis blade profile & $5.21{\times}10^{-5}$ $\mathrm{s}^{2}/\mathrm{m}^{2}$\\
        \hline
        $C_{2}$ & UAV power constant vis-\'{a}-vis induced effects & $709.27$ W\\
        \hline
        $C_{3}$ & UAV power constant vis-\'{a}-vis induced effects & $129.92$ $\mathrm{s}^{2}/\mathrm{m}^{2}$\\
        \hline
        $C_{4}$ & UAV power constant vis-\'{a}-vis parasitic effects & $0.02$ W\\
        \hline
        $g$; $\rho$ & Acceleration under gravity; Air density & $9.81$ m/$\mathrm{s}^{2}$; $1.23$ kg/$\mathrm{m}^{3}$\\
        \hline
        $\varphi$; $\vartheta$ & UAV rotor solidity; UAV rotor disc area & $0.1$; $0.5$ $\mathrm{m}^{2}$\\
        \hline
        $\omega$; $\varrho$ & UAV fuselage drag ratio; UAV weight & $0.6$; $80$ N\\
        \hline
        $v_{\mathrm{max}}$; $a_{\mathrm{max}}$ & UAV max velocity; UAV max acceleration & $50$ m/s; $5$ m/$\mathrm{s}^{2}$\\
        \hline
        \hline
        $N_{\mathrm{sw}}$; $N_{\mathrm{ssw}}$ & LCSO swarm size; LCSO sub-swarm size & $180$; $20$\\
        \hline
        $M_{\mathrm{seg}}$; $F_{\mathrm{max}}$ & LCSO segment size; LCSO max evaluations count & $128$; $1000$\\
        \hline
        \end{tabular}
        \vspace{-1mm}
        \caption{The simulation parameters employed in our numerical evaluations.}
        \vspace{-3mm}
        \label{T2}
    \end{center}
\end{table}
The objective of the proposed solution framework (Fig.~\ref{F3}) is to maximize the cumulative fleet-wide reward obtained by successfully harvesting prioritized traffic from GNs on-site over a preset finite mission duration. Let $\mathcal{S}$ represent the set of optimization variables, i.e., the beam-forming design, the optimal UAV serving positions, the $3$D energy-conscious UAV trajectories, and the GN association/scheduling policy. Then, the optimization problem is described as follows in Eq.~\eqref{Optimization_problem}, with the constraints including the UAV start and end position enforcement in \eqref{C1}; the UAV collision avoidance in \eqref{C2}; the GN-UAV scheduling exclusivity in \eqref{C3}; and the UAV's average mobility power consumption, the UAV's velocity bounds, and the UAV's acceleration bounds, in \eqref{C4}:
\vspace{-5mm}
\begin{align}\label{Optimization_problem}
    &\underset{\mathcal{S}}{\mathrm{maximize}}\sum_{t{=}1}^{T}\sum_{u{\in}\mathcal{U}}\sum_{g{\in}\mathcal{G}_{u}}\Omega_{gu}\text{ s.t. } \forall u \in \mathcal{U},\  \forall t \in [0,T],\\
    &\mathbf{p}_{u}(0) \in \mathcal{L},\ \mathbf{p}_{u}(T) \in \mathcal{L},\label{C1}\tag{C.1}\\
    &\mathbf{p}_{u_{1}}(t) \neq \mathbf{p}_{u_{2}}(t),\ \forall u_{1}, u_{2} \in \mathcal{U},\ u_{1} \neq u_{2},\label{C2}\tag{C.2}\\
    &g \in \mathcal{G}_{u_{1}}{\implies}g \notin \mathcal{G}_{u_{2}}, u_{1}, u_{2} \in \mathcal{U}, u_{1} \neq u_{2}, g \in \mathcal{G},\label{C3}\tag{C.3}\\
    &P_{u,3\mathrm{D}}(t) \leq P_{\mathrm{avg}},\ \abs{\vec{v}_{u}(t)} \leq v_{\mathrm{max}},\ \abs{\vec{a}_{u}(t)} \leq a_{\mathrm{max}}.\label{C4}\tag{C.4}
\end{align}
Given the problem complexity, we decompose it into radio and vehicle layer sub-problems, and solve each individually using its corresponding proposed algorithm (Fig.~\ref{F3}). First, we cluster the GNs according to their proximity on-site using the naive (standard) K-means clustering algorithm; then, coupled with zero-forcing beam-forming design, we employ a coarse- and fine-grained grid search to find the optimal UAV service voxel; next, using a learning-based competitive swarm optimization algorithm, we design the energy-conscious UAV trajectories; finally, with a multiple traveling salesman problem setup, we solve for the GN association/scheduling policy via a graphical branch-and-bound technique. We discuss these in detail next.\\
\noindent{\textbf{Radio Layer | Optimal UAV Positioning}}: Upon clustering the GNs on-site into $C$ clusters, we employ two-stage grid search to determine the optimal $3$D positioning of a UAV serving each of these GN clusters. The first stage (coarse search) involves a bounding-box strategy to determine the larger set of $3$D grid voxels, i.e., the voxels within the smallest area subset encapsulating the GNs in a cluster; these constitute the argument set for the second stage (fine search), where with Zero-Forcing (ZF) beam-forming to design the precoding and combining matrices described in Sec.~\ref{S2} according to the procedures outlined in~\cite{ZF, Core_SoA_1}, we select the grid voxel that maximizes the cluster reward as the optimal service position.\\
\noindent{\textbf{Vehicle Layer | UAV Trajectory Design}}: The Learning based Competitive Swarm Optimization (LCSO) algorithm~\cite{LCSO} is a computationally efficient variant of the popular Competitive Swarm Optimization (CSO) algorithm typically used for route planning~\cite{MAESTRO_TCCN}, thereby making it a suitable candidate for $3$D UAV trajectory design. Treating the optimal service positions for each GN cluster as the graph vertices of a fully-connected mesh (including the depot), we use LCSO to obtain the $3$D trajectory a UAV in the fleet should execute vis-\'{a}-vis an edge connecting any two of these graph vertices. In this trajectory design process, a Lagrangian setup involving the average mobility power consumption constraint imposed on each UAV is introduced, wherein we use projected subgradient ascent (iteratively coupled with LCSO) for dual variable optimization. First, the randomly initialized swarm of $N_{\mathrm{sw}}$ particles $(\mathbf{p},v)_{1:N_{\mathrm{sw}}}$ and their particle velocities $(\boldsymbol{\upsilon},\eta)_{1:N_{\mathrm{sw}}}$ is grouped into sub-swarms of $N_{\mathrm{ssw}}$ particles; then, in each iteration $i$, every particle $(\mathbf{p}, v)$ in a sub-swarm competes in a tournament with others in the sub-swarm as follows based on a mobility power cost function (see Sec.~\ref{S2}). These tournaments are organized independently (and in parallel) across all the sub-swarms. Specifically, each such tournament within any sub-swarm involves the following procedure: first, randomly select three particles; next, compute and compare their power cost functions; consequently, decide a winner $\boldsymbol{\psi}_{w}$, a runner-up $\boldsymbol{\psi}_{r}$, and a loser $\boldsymbol{\psi}_{l}$; subsequently, modify the knowledge of the runner-up and the loser based on the winner as follows:
\vspace{-2.4mm}
\begin{align}\label{LCSO}
    &\boldsymbol{\xi}_{r}(i + 1) = n_{1}\boldsymbol{\xi}_{r}(i) + n_{2}\left(\boldsymbol{\psi}_{w}(i) - \boldsymbol{\psi}_{r}(i)\right);\nonumber\\
    &\boldsymbol{\psi}_{r}(i + 1) = \boldsymbol{\psi}_{r}(i) + \boldsymbol{\xi}_{r}(i + 1);\nonumber\\
    &\boldsymbol{\xi}_{l}(i{+}1){=}n_{1}\boldsymbol{\xi}_{l}(i){+}n_{2}\left(\boldsymbol{\psi}_{w}(i){-}\boldsymbol{\psi}_{l}(i)\right){+}n_{3}\left(\boldsymbol{\psi}_{r}(i){-}\boldsymbol{\psi}_{l}(i)\right);\nonumber\\
    &\boldsymbol{\psi}_{l}(i + 1) = \boldsymbol{\psi}_{l}(i) + \boldsymbol{\xi}_{l}(i + 1);
\end{align}
where $\boldsymbol{\xi}_{w}$, $\boldsymbol{\xi}_{r}$, and $\boldsymbol{\xi}_{l}$ correspond to the particle velocities of the winner, runner-up, and loser particles, $\boldsymbol{\psi}_{w}$, $\boldsymbol{\psi}_{r}$, and $\boldsymbol{\psi}_{l}$, respectively; and $n_{1},n_{2},n_{3}{\sim}\mathrm{Uniform}[0,1]$. Next, the second stage of the LCSO algorithm constitutes a tournament among the sub-swarms: randomly choose a particle from among the winners of each sub-swarm; then, randomly choose three particles from this set of winners; after which, compute and compare their power cost functions to determine a winner $\boldsymbol{\psi}_{w}$, a runner-up $\boldsymbol{\psi}_{r}$, and a loser $\boldsymbol{\psi}_{l}$; next, update the knowledge of the runner-up and the loser particles based on the attributes of the winner according to~\eqref{LCSO}. This iterative two-stage process continues until the number of constituent computations of the underlying power costs exceeds a preset threshold ($F_{\mathrm{max}}$)~\cite{LCSO}.
\noindent{\textbf{Radio + Vehicle Layers | GN Association/Scheduling}}: To solve for the GN association/scheduling policy, we formulate a multiple Traveling Salesman Problem (mTSP) setup~\cite{mTSP_Branch_and_Bound}, wherein with the added constraints of the UAVs having to start and end at the depot, the inherent objective is to plan the routes of the UAVs in the network to serve the GN clusters on-site (at the corresponding optimal service voxels obtained earlier via two-stage grid search), while maximizing the cumulative reward attained across the fleet over the mission execution duration. In the overlaid fully-connected mesh, with the graph vertices (i.e., the optimal service positions and the depot) and the edges connecting them (i.e., the $3$D UAV trajectories and their time \& power costs) obtained via the processes outlined earlier in this section, this mTSP formulation is solved using the well-known graphical branch-and-bound method~\cite{mTSP_Branch_and_Bound} to obtain the association of UAVs in the fleet with specific GN clusters as well as the scheduling sequence in which a particular UAV goes about serving its associated clusters.
\vspace{-3mm}

% Numerical evaluations
\section{Numerical Evaluations}\label{S4}
\begin{figure*}[t]
    \centering
    \begin{subfigure}{0.494\linewidth}
        \centering
        \includegraphics[width=0.71\linewidth]{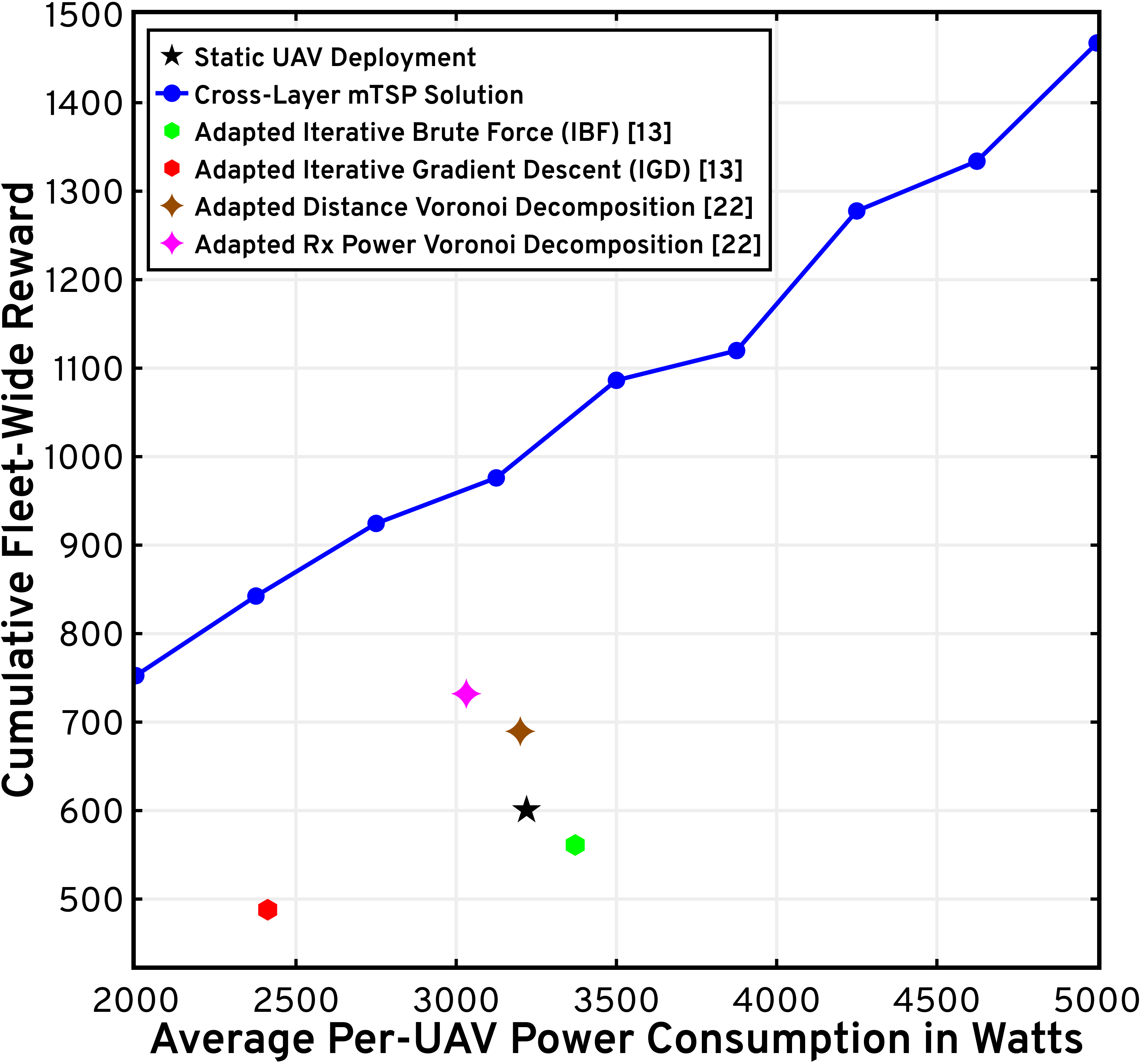}
        \vspace{-1.5mm}
        \caption{Cumulative Fleet-Wide Reward vs Average Per-UAV Power Consumption~\cite{Source_Code}}
        \label{F4a}
    \end{subfigure}
    \begin{subfigure}{0.496\linewidth}
        \centering
        \includegraphics[width=0.71\linewidth]{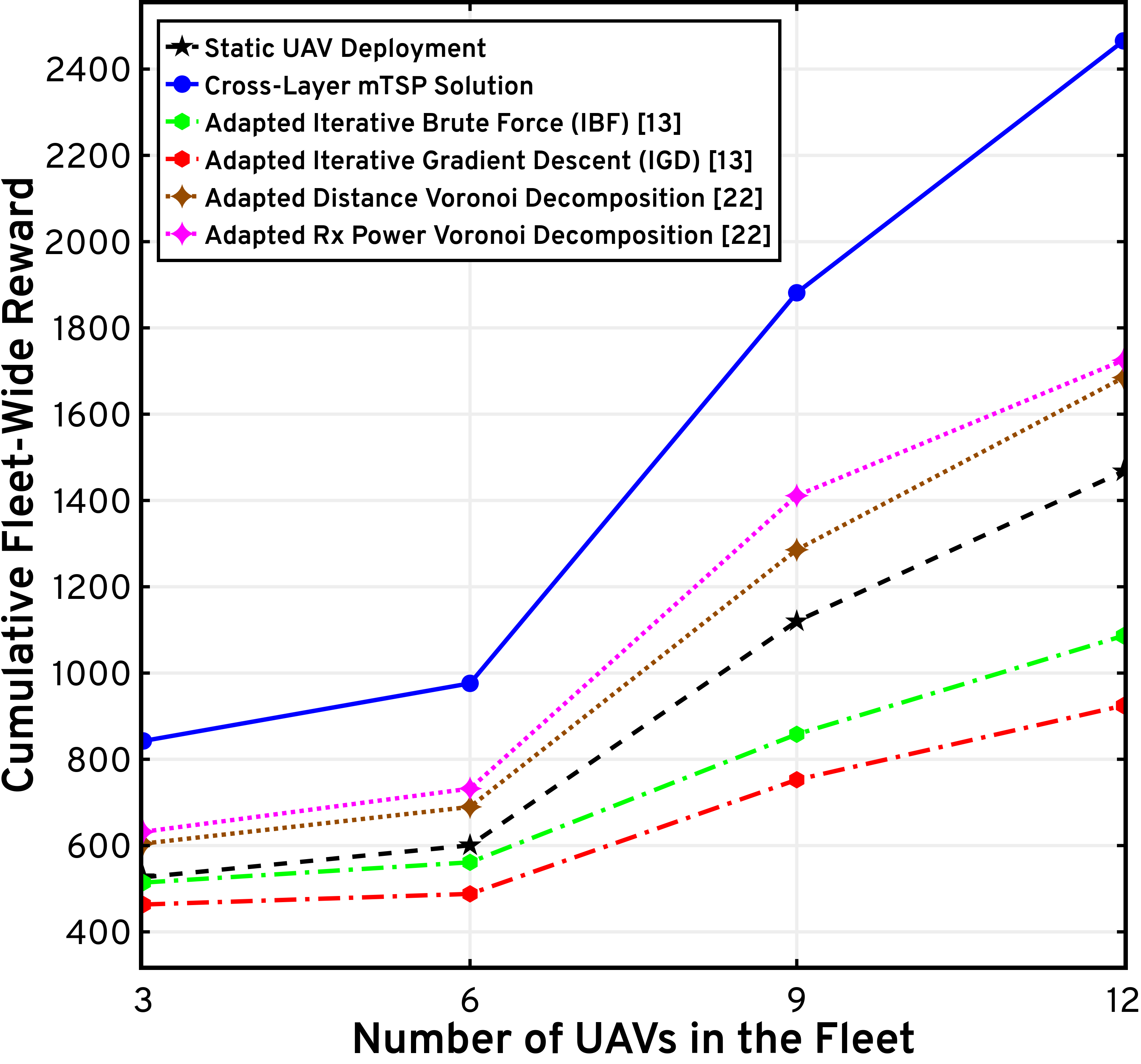}
        \vspace{-1.5mm}
        \caption{Cumulative Fleet-Wide Reward vs Number of UAVs in the Fleet~\cite{Source_Code}}
        \label{F4b}
    \end{subfigure}
    \vspace{-6mm}
    \caption{(a) A plot of the total fleet-wide reward vs average per-UAV power consumption; (b) A plot of the total fleet-wide reward vs the number of UAVs.}
    \vspace{-7mm}
    \label{F4}
\end{figure*}
With the simulation setup detailed in Table~\ref{T2}, we evaluate the performance of the proposed cross-layer optimization framework against static UAV deployments, adaptive Voronoi decompositions~\cite{Voronoi}, and the iterative fleet orchestration schemes adapted from~\cite{CORES_ICASSP}. For static UAV deployments, coupled with K-means clustering ($C{=}U{=}6$) and a multi-user MIMO ZF beam-forming design, we statically position each of the $U$ UAVs at the centroid of their respective clusters (at a height of $z_{u}{=}145$ m,${\forall}u{\in}\mathcal{U}$) and evaluate the fleet-wide reward. For the adaptive Voronoi decomposition techniques, we iteratively update (until convergence) the Voronoi sets of associated GNs for each UAV, using either the GN-UAV distance or the received power as the evaluation metric; then, we update the UAV positions to be centroids of their respective Voronoi sets in that iteration; finally, we position the UAVs at these optimal positions and compute the fleet-wide reward. For the Iterative Gradient Descent (IGD) and the Iterative Brute Force (IBF) algorithms adapted from~\cite{CORES_ICASSP} to suit our network modeling, upon clustering the GNs via K-means clustering ($C{=}U{=}6$), coupled with ZF beam-forming, the optimal UAV service position for each GN cluster is obtained via the IGD and IBF algorithms~\cite{CORES_ICASSP}; subsequently, we positions the UAVs at these optimal service positions and evaluate the fleet-wide reward. Fig.~\ref{F4a} depicts the total reward accumulated by the fleet of UAVs using our solution, as a function of the average power consumption constraint on each UAV ($P_{\mathrm{avg}}$), over the simulated mission duration ($T{=}3000$ s). For similar power levels, we demonstrate a $39$\% gain over static deployments, a $31$\% gain over distance based Voronoi decomposition~\cite{Voronoi}, a $23$\% gain over Rx power based Voronoi decomposition~\cite{Voronoi}, a $42$\% gain over the IGD scheme from~\cite{CORES_ICASSP}, and a $47$\% gain over the IBF scheme from~\cite{CORES_ICASSP}. Also, Fig.~\ref{F4b} shows the cumulative fleet-wide reward as a function of the number of UAVs $U$ in the fleet (fixed $G{=}36$) along with the reference benchmarks of a static deployment, the Voronoi decomposition techniques~\cite{Voronoi}, and the IGD and IBF schemes from~\cite{CORES_ICASSP}: here, we show that, the cumulative fleet-wide reward increases with the number of UAVs, and that our proposed cross-layer mTSP solution consistently (across $3$, $6$, $9$, and $12$ UAVs) outperforms static UAV deployments, adaptive Voronoi decompositions~\cite{Voronoi}, and IGD and IBF schemes from~\cite{CORES_ICASSP}. 
\vspace{-4mm}

% Concluding remarks and Future work
\section{Conclusion}\label{S5}
We detail the orchestration of a fleet of MIMO-capable rotary-wing UAVs for prioritized data harvesting from GNs (with MIMO capabilities). With a preset mission duration and site tessellation, the fleet-wide reward maximization problem is solved offline under a cross-layer optimization formulation. With K-means clustering and ZF beam-forming, we employ two-stage grid search to obtain the optimal UAV service voxel; next, we design the energy-conscious $3$D UAV trajectories via LCSO; finally, we derive the GN association/scheduling policy via a branch-and-bound method (which solves the underlying mTSP). Numerical evaluations illustrate that our solution framework outperforms static UAV deployments, adaptive Voronoi decompositions, as well as the IGD and IBF schemes vis-\'{a}-vis GN quality-of-service and UAV power efficiency.
\vspace{-4mm}

% References (main.bib)
\bibliographystyle{IEEEtran}
\bibliography{IEEEabrv,main}

\begin{thebibliography}{10}
\providecommand{\url}[1]{#1}
\csname url@samestyle\endcsname
\providecommand{\newblock}{\relax}
\providecommand{\bibinfo}[2]{#2}
\providecommand{\BIBentrySTDinterwordspacing}{\spaceskip=0pt\relax}
\providecommand{\BIBentryALTinterwordstretchfactor}{4}
\providecommand{\BIBentryALTinterwordspacing}{\spaceskip=\fontdimen2\font plus
\BIBentryALTinterwordstretchfactor\fontdimen3\font minus \fontdimen4\font\relax}
\providecommand{\BIBforeignlanguage}[2]{{%
\expandafter\ifx\csname l@#1\endcsname\relax
\typeout{** WARNING: IEEEtran.bst: No hyphenation pattern has been}%
\typeout{** loaded for the language `#1'. Using the pattern for}%
\typeout{** the default language instead.}%
\else
\language=\csname l@#1\endcsname
\fi
#2}}
\providecommand{\BIBdecl}{\relax}
\BIBdecl

\bibitem{Source_Code}
\BIBentryALTinterwordspacing
B.~Keshavamurthy, ``{ACCUSTOM: Adaptive Control and Coordination of UAV Swarms for Traffic Offloading in MIMO ecosystems}.'' [Online]. Available: \url{https://github.com/bharathkeshavamurthy/ACCUSTOM.git}
\BIBentrySTDinterwordspacing

\bibitem{Motivation_1}
B.~Yang, E.~Yang \emph{et~al.}, ``{Ultrasonic- and IMU-Based High-Precision UAV Localization for the Low-Cost Autonomous Inspection in Oil and Gas Pressure Vessels},'' \emph{IEEE Trans. Ind. Informat.}, vol.~19, no.~10, pp. 10\,523--10\,534, 2023.

\bibitem{Motivation_2}
H.~M. Chung, S.~Maharjan \emph{et~al.}, ``{Placement and Routing Optimization for Automated Inspection With Unmanned Aerial Vehicles: A Study in Offshore Wind Farm},'' \emph{IEEE Trans. Ind. Informat.}, vol.~17, no.~5, pp. 3032--3043, 2021.

\bibitem{MAESTRO_TCCN}
B.~Keshavamurthy, M.~A. Bliss \emph{et~al.}, ``{MAESTRO-X: Distributed Orchestration of Rotary-Wing UAV-Relay Swarms},'' \emph{IEEE Trans. Cogn. Commun. Netw.}, vol.~9, no.~3, pp. 794--810, 2023.

\bibitem{SPAVE_ICC}
B.~Keshavamurthy, Y.~Zhang \emph{et~al.}, ``{Propagation Measurements and Analyses at 28 GHz via an Autonomous Beam-Steering Platform},'' in \emph{2023 IEEE Intl. Conf. Commun.}, 2023, pp. 5042--5047.

\bibitem{Core_SoA_1_Ref_13}
J.~Lyu, Y.~Zeng \emph{et~al.}, ``{Placement Optimization of UAV-Mounted Mobile Base Stations},'' \emph{IEEE Commun. Lett.}, vol.~21, no.~3, pp. 604--607, 2017.

\bibitem{Core_SoA_1_Ref_14}
M.~Mozaffari, W.~Saad \emph{et~al.}, ``{Efficient Deployment of Multiple Unmanned Aerial Vehicles for Optimal Wireless Coverage},'' \emph{IEEE Commun. Lett.}, vol.~20, no.~8, pp. 1647--1650, 2016.

\bibitem{Core_SoA_1_Ref_12}
M.~Mozaffari, W.~Saad \emph{et~al.}, ``{Mobile Unmanned Aerial Vehicles (UAVs) for Energy-Efficient Internet of Things Communications},'' \emph{IEEE Trans. Wireless Commun.}, vol.~16, no.~11, pp. 7574--7589, 2017.

\bibitem{Core_SoA_1_Ref_18_Extended_From_17}
M.~Hua, L.~Yang \emph{et~al.}, ``{3D UAV Trajectory and Communication Design for Simultaneous Uplink and Downlink Transmission},'' \emph{IEEE Trans. Commun.}, vol.~68, no.~9, pp. 5908--5923, 2020.

\bibitem{Core_SoA_1_Ref_25}
X.~Pang, J.~Tang \emph{et~al.}, ``{Energy-efficient design for mmWave-enabled NOMA-UAV networks},'' \emph{Sci. China Inf. Sci.}, vol.~64, no.~4, Apr 2021.

\bibitem{Core_SoA_1}
N.~Nouri, F.~Fazel \emph{et~al.}, ``{Multi-UAV Placement and User Association in Uplink MIMO Ultra-Dense Wireless Networks},'' \emph{IEEE Trans. Mob. Comput.}, vol.~22, no.~3, pp. 1615--1632, 2023.

\bibitem{Core_SoA_1_Ref_24}
N.~Nouri, J.~Abouei \emph{et~al.}, ``{Three-Dimensional Multi-UAV Placement and Resource Allocation for Energy-Efficient IoT Communication},'' \emph{IEEE Internet Things J.}, vol.~9, no.~3, pp. 2134--2152, 2022.

\bibitem{CORES_ICASSP}
S.~Hanna, H.~Yan \emph{et~al.}, ``{Distributed UAV Placement Optimization for Cooperative Line-of-sight MIMO Communications},'' in \emph{2019 Proc. IEEE Int. Conf. Acoust. Speech Signal Process.}, 2019, pp. 4619--4623.

\bibitem{CORES_JSAC}
S.~Hanna, E.~Krijestorac \emph{et~al.}, ``{UAV Swarm Position Optimization for High Capacity MIMO Backhaul},'' \emph{IEEE J. Sel. Areas Commun.}, vol.~39, no.~10, pp. 3006--3021, 2021.

\bibitem{Core_SoA_1_Ref_27_Related_To_26}
W.~Feng, N.~Zhao \emph{et~al.}, ``{Joint 3D Trajectory Design and Time Allocation for UAV-Enabled Wireless Power Transfer Networks},'' \emph{IEEE Trans. Veh. Technol.}, vol.~69, no.~9, pp. 9265--9278, 2020.

\bibitem{DARPA:SC2}
M.~Rosker, ``{Spectrum Collaboration Challenge (SC2)},'' \emph{DARPA Spectrum Collaboration Challenge (SC2)}, 2018.

\bibitem{UAV_Propulsion_2}
N.~Gao, Y.~Zeng \emph{et~al.}, ``{Energy model for UAV communications: Experimental validation and model generalization},'' \emph{China Communications}, vol.~18, no.~7, pp. 253--264, 2021.

\bibitem{UAV_Propulsion_1}
H.~Yan, Y.~Chen \emph{et~al.}, ``{New Energy Consumption Model for Rotary-Wing UAV Propulsion},'' \emph{IEEE Wireless Commun. Lett.}, vol.~10, no.~9, pp. 2009--2012, 2021.

\bibitem{ZF}
N.~Jindal, ``{MIMO Broadcast Channels With Finite-Rate Feedback},'' \emph{IEEE Trans. Inf. Theory}, vol.~52, no.~11, pp. 5045--5060, 2006.

\bibitem{LCSO}
B.~Borowska, ``{Learning Competitive Swarm Optimization},'' \emph{Entropy}, vol.~24, no.~2, 2022.

\bibitem{mTSP_Branch_and_Bound}
B.~Gavish and K.~Srikanth, ``{An Optimal Solution Method for Large-Scale Multiple Traveling Salesmen Problems},'' \emph{Operations Research}, vol.~34, no.~5, pp. 698--717, 1986.

\bibitem{Voronoi}
M.~E. Morocho-Cayamcela, W.~Lim \emph{et~al.}, ``{An optimal location strategy for multiple drone base stations in massive MIMO},'' \emph{ICT Express}, vol.~8, no.~2, pp. 230--234, 2022.

\end{thebibliography}

\end{document}